\documentclass[a4paper,11pt]{article}
\usepackage{pos}

\usepackage{subfigure}

\title{Search for dark sector at BESIII}

\author*[a]{Zhi-Jun Li}
\author[a]{Zheng-Yun You}

\affiliation[a]{School of Physics, Sun Yat-sen University, Guangzhou 510275, China}

\emailAdd{lizhj37@mail2.sysu.edu.cn}
\emailAdd{youzhy5@mail.sysu.edu.cn}

\abstract{BESIII experiment has collected a large data sample of charmonium, charm mesons, hyperons, and other light mesons. These data provide a unique opportunity to explore the dark sector beyond the Standard Model, particularly for dark sectors that couple to charm quarks or other light quarks, and for dark sectors with masses in the $\tau$–c energy region.
We present recent dark sector search results from the BESIII experiment, including the search for massless dark photons in $D^0\to\omega\gamma'$ and $D^0\to\gamma\gamma'$, the search for invisible decays of $K^0_S$, the search for dark baryon particles in $\Xi^-\to\pi^-\chi$, the search for massless particles in $\Sigma^+\to p + \text{invisible}$, and the search for axion-like particles in $J/\psi\to\gamma a$ with $a\to\gamma\gamma$.}

\FullConference{The European Physical Society Conference on High Energy Physics (EPS-HEP2025)\\
7-11 July 2025\\
Marseille, France\\}

\begin{document}
\maketitle

\section{Introduction}
The Standard Model (SM) has achieved significant success in explaining most phenomena in our universe, encompassing both fundamental particles and their interactions. However, several unresolved puzzles remain, including dark matter (DM), the strong CP problem, the muon $g-2$ anomaly, the fermion mass hierarchy, matter-antimatter asymmetry, and others. These challenges suggest the existence of a dark sector beyond the SM, potentially containing new particles and interactions between the dark sector and SM matter.
The search for the dark sector provides an intriguing avenue for probing new physics (NP) beyond the SM. If the masses of dark sector particles fall within the 0 to GeV range, they can be accessed through high-intensity $e^+e^-$ collider experiments such as the Beijing Spectrometer III (BESIII)~\cite{BESIII:2009fln}.

BESIII is a general-purpose spectrometer designed for investigating $\tau$-charm physics within a center-of-mass energy from 2.0 to 4.7~GeV. It records symmetric electron-positron ($e^+e^-$) collisions generated by the Beijing Electron Positron Collider II (BEPCII) storage ring~\cite{Yu:2016cof}. To date, BESIII has collected extensive datasets in this energy range, including 10 billion $J/\psi$ events, 2.7 billion $\psi(2S)$ events, a 20~fb$^{-1}$ dataset at 3.773 GeV, and upwards of 20~fb$^{-1}$ of data above 4.0 GeV~\cite{Liao:2025lth}. These large charmonium samples also enable the production of significant samples of hyperons and light mesons through charmonium decay processes. By capitalizing on these extensive datasets and utilizing advanced analytical techniques~\cite{Li:2024pox,Song:2025pnt}, BESIII is well-positioned to conduct in-depth investigations into the dark sector beyond the SM.

\section{Search for massless dark photon with $D^0\to\omega\gamma'$ and $D^0\to\gamma\gamma'$}
The dark photon ($\gamma'$) is a hypothetical vector boson introduced in minimal extensions of the SM through an additional Abelian gauge group, which has a kinetic mixing with the SM photon. When the symmetry of this additional Abelian gauge group is spontaneously broken, the dark photon acquires mass and directly couples to SM fermions. Conversely, if the symmetry remains unbroken, the dark photon remains massless and does not directly couple to SM particles from a theoretical standpoint. 
The massless dark photon plays a crucial role in the dark sector by providing a new long-range force for DM, offering a potential solution to the fermion mass hierarchy problem, explaining the excess observed in the decay process $B^+ \to K^+ \nu\bar{\nu}$, elucidating the origin of the CKM matrix structure, and addressing the vacuum instability issue in the SM Higgs sector~\cite{Fabbrichesi:2020wbt}.

In contrast to the stringent constraints placed on massive dark photons, the massless dark photon remains considerably less restricted. Searches for massless dark photons need to rely on higher-dimensional operators, such as~\cite{Dobrescu:2004wz}
\begin{eqnarray}
\mathcal{L}_{\rm{NP}} = & \frac{1}{\Lambda^2_{\rm{NP}}} \left( C^u_{jk} \bar{q}_j \sigma^{\mu\nu} u_k \tilde{H} + C^d_{jk} \bar{q}_j \sigma^{\mu\nu} d_k H + C^l_{jk} \bar{l}_j \sigma^{\mu\nu} e_k H + h.c. \right) F'_{\mu\nu}.
\label{eq:dimension-six operator}
\end{eqnarray}
Here, $\Lambda_{\rm{NP}}$ represents the NP energy scale, and $C_{jk}$ are dimensionless coefficients. This operator inherently involves flavor-violating interactions, such as the coupling between charm and up quarks with the dark photon ($cu \gamma'$).
Utilizing $7.9~\rm{fb}^{-1}$ of $e^+e^-$ annihilation data at a center-of-mass energy of $\sqrt{s} = 3.773$ GeV, we search for the massless dark photon in the decay channels $D^0 \to \omega \gamma'$ and $D^0 \to \gamma \gamma'$. At this energy, $D^0$ mesons are produced in pairs, allowing one $D^0$ to be tagged using well-established SM hadronic decay modes. We then investigate the NP decays on the opposite side of the tagged $D^0$, employing a technique known as the double tag method.

\begin{figure*}[htbp] \centering
	\setlength{\abovecaptionskip}{-1pt}
	\setlength{\belowcaptionskip}{10pt}

        \subfigure[]
        {\includegraphics[width=0.32\textwidth]{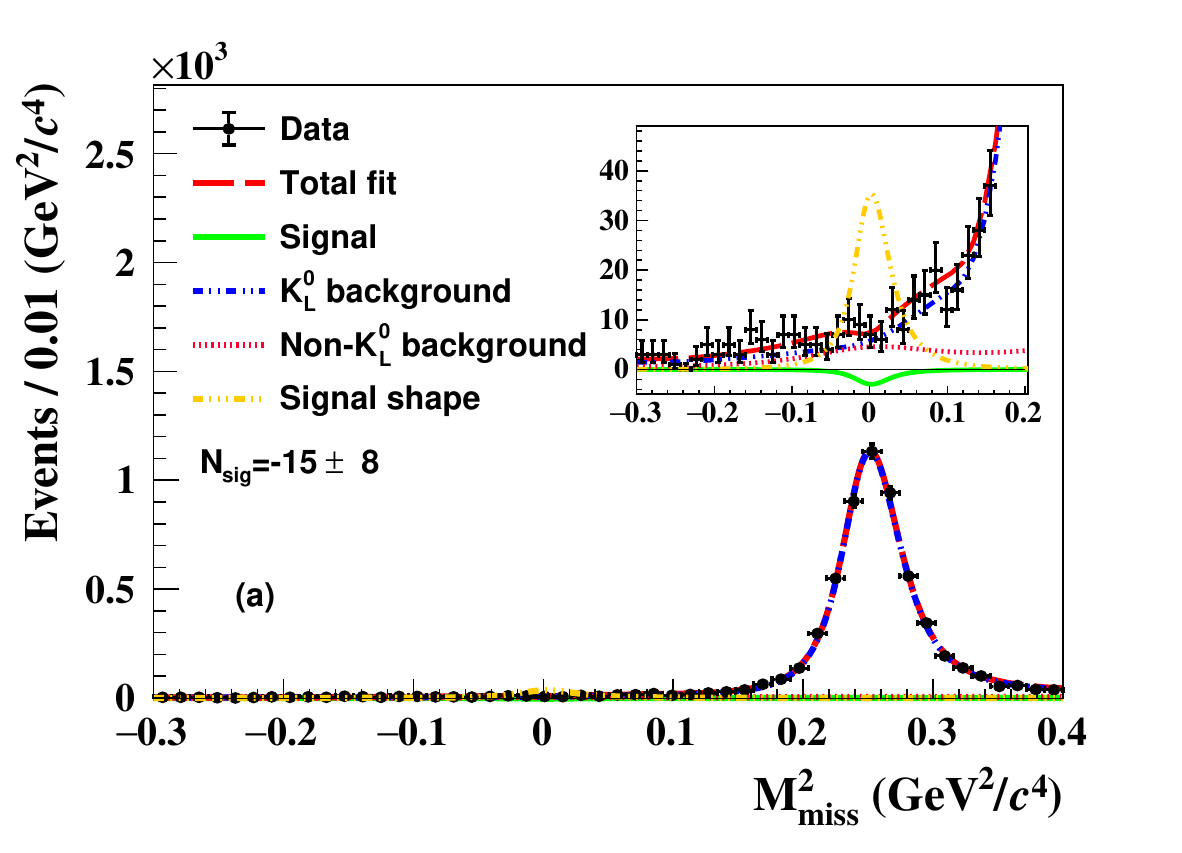}}
        \subfigure[]
        {\includegraphics[width=0.32\textwidth]{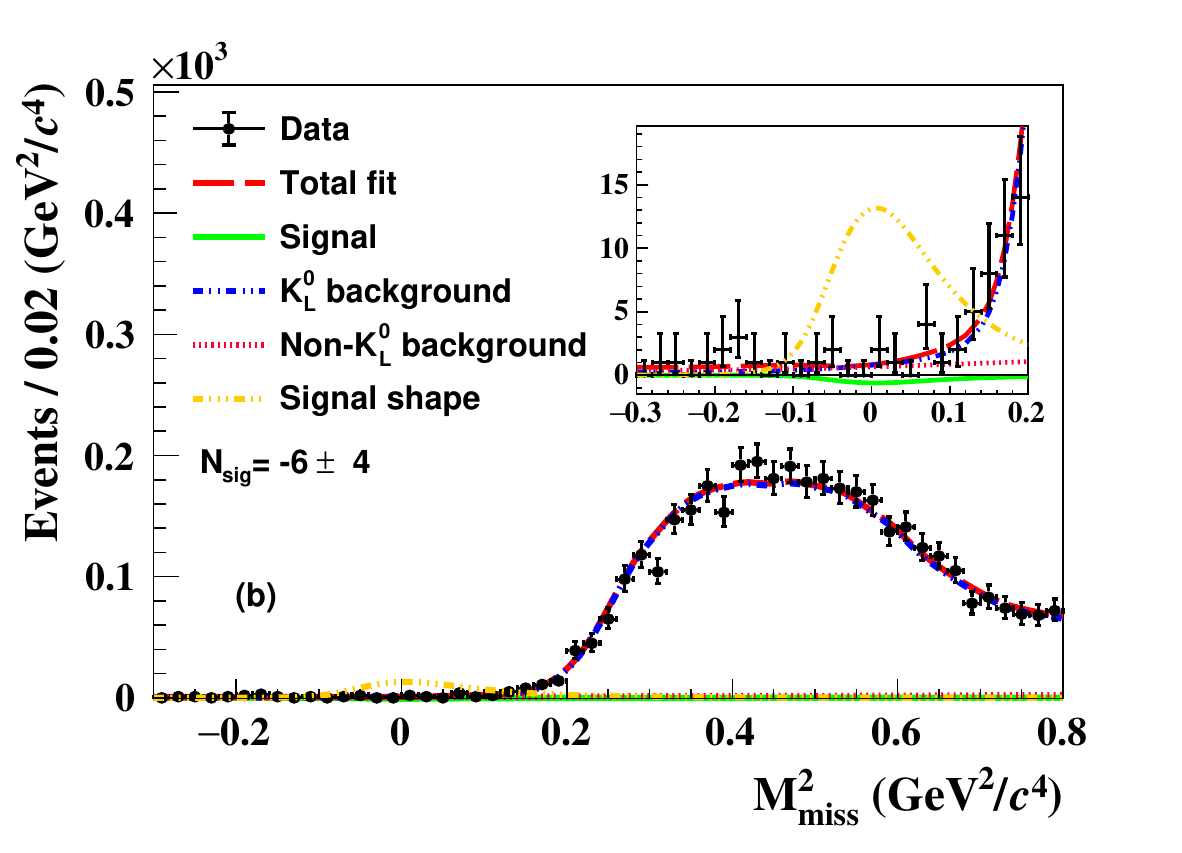}}
        \subfigure[]
        {\includegraphics[width=0.32\textwidth]{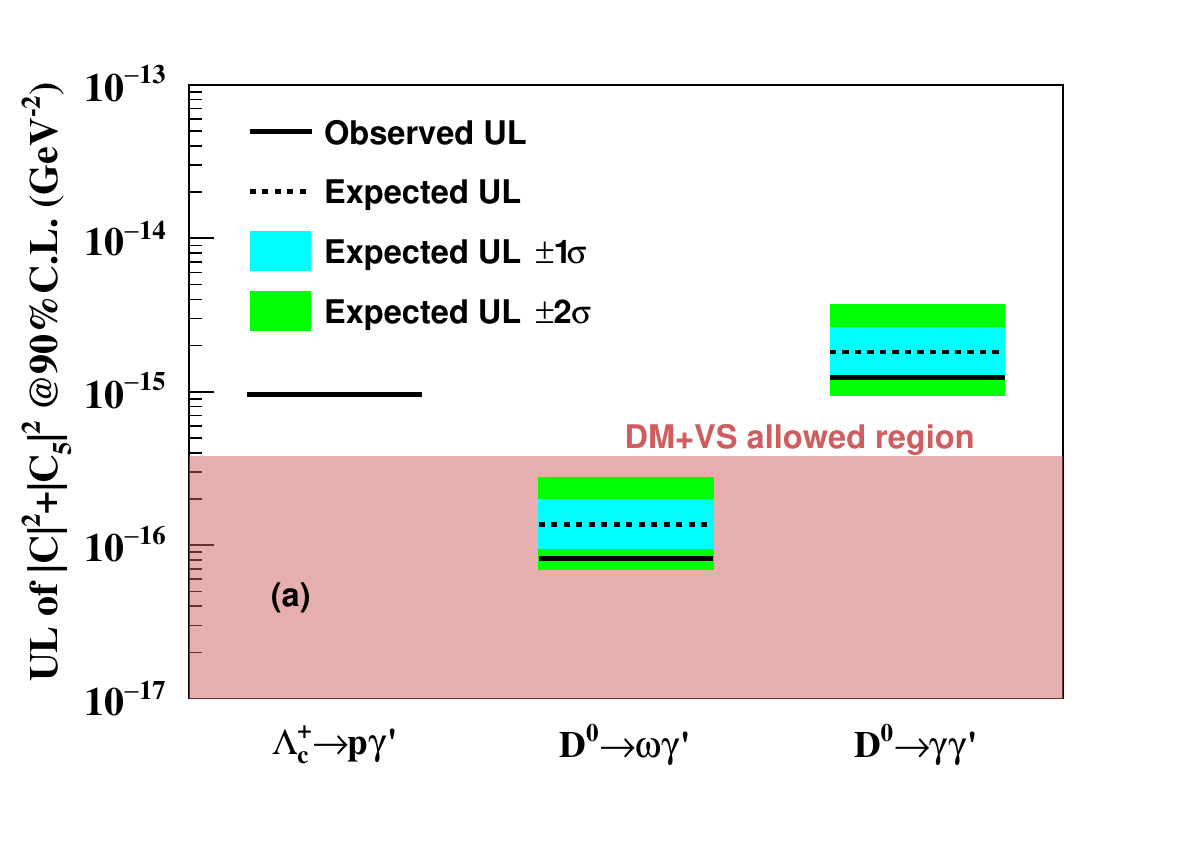}}\\
        
	\caption{
        (a) and (b) display the $M^2_{\rm{miss}}$ distributions for the $D^0 \to \omega \gamma'$ and $D^0 \to \gamma \gamma'$ decay candidates, respectively. The black points represent the observed data, while the other lines correspond to various background and signal components. (c) illustrates the constraints on the $cu \gamma'$ coupling. The black solid lines indicate the observed ULs from different processes, and the red shaded regions denote the allowed parameter space from DM and VS.
        } 
	\label{fig:Gp}
\end{figure*}

The invisible massless dark photon can be inferred from the recoil of visible SM particles (tagged $D^0$ and $\omega$ or $\gamma$), as illustrated in Figure~\ref{fig:Gp} (a) and (b). No significant signals are observed in the data samples. The upper limits (ULs) on the branching fraction (BF) at the 90\% confidence level (C.L.) are determined to be $1.1 \times 10^{-5}$ and $2.0 \times 10^{-6}$ for the decay processes $D^0 \to \omega \gamma'$ and $D^0 \to \gamma \gamma'$, respectively~\cite{BESIII:2024rkp}.
\vspace{-0.0cm}
\vspace{-0.0cm}
The BF of $D^0$ decaying to a massless dark photon is highly sensitive to the NP energy scale and the dimensionless coefficients introduced in Eq.~\ref{eq:dimension-six operator}, with the relationship given by~\cite{Su:2020yze} of 
$\mathcal{B}(D^0 \to V \gamma') = \frac{\tau_D f^2_{DV} (m^2_D - m^2_V)^3}{2\pi m^3_D} (|\mathcal{C}|^2 + |\mathcal{C}_5|^2)$,
and 
$\mathcal{B}(D^0 \to \gamma \gamma') = \frac{\alpha_e}{2} \tau_D f^2_{D\gamma} m^3_{D} (|\mathcal{C}|^2 + |\mathcal{C}_5|^2)$,
where $f$ denotes the form factor associated with $D$ meson decays, $\mathcal{C} = \Lambda_{\rm{NP}}^{-2} \left( C^u_{12} + C^{u*}_{12} \right) \nu / \sqrt{8}$ and $\mathcal{C}_5 = \Lambda_{\rm{NP}}^{-2} \left( C^u_{12} - C^{u*}_{12} \right) \nu / \sqrt{8}$, and $\nu$ represents the vacuum expectation value of the Higgs field.
For the decay channel $D^0 \to \omega \gamma'$, we establish a stringent constraint on the parameters related to the NP energy scale, specifically $|\mathcal{C}|^2 + |\mathcal{C}_5|^2 < 8.2 \times 10^{-17}~\rm{GeV}^{-2}$. This constraint signifies an improvement of more than one order of magnitude compared to previous result from $\Lambda_c\to p\gamma'$ and reaches in the allowed regions from DM and vacuum stability (VS) for the first time, as illustrated in Figure~\ref{fig:Gp} (c). Conversely, the constraint derived from the decay $D^0 \to \gamma \gamma'$, despite presenting a more stringent UL on the BF, is comparatively weaker due to the additional factor of $\alpha_e$ involved in the decay process.

\section{Search for the invisible decay of $K^0_S$}
The BF for the invisible decay of $K^0_S$ is inherently very small in the SM, primarily due to loop-level diagrams and helicity suppression. However, certain NP scenarios beyond the SM could potentially enhance this BF. For example, if $K^0_S$ decays into a pair of DM particles or undergoes ordinary-mirror particle oscillations, the invisible BF could be increased to approximately $10^{-6}$. 
Moreover, the invisible decay of $K^0_S$ provides crucial insights for testing CPT (Charge, Parity, and Time reversal) symmetry. The Bell-Steinberger relation, which connects CPT violation to the amplitudes of all decay channels of neutral kaons, relies on the current assumption that no invisible decay modes exist~\cite{Gninenko:2014sxa}. Therefore, observing or constraining the invisible decay modes of $K^0_S$ can offer valuable information for validating CPT symmetry.

\vspace{-0.0cm}
\begin{figure*}[htbp] \centering
	\setlength{\abovecaptionskip}{-1pt}
	\setlength{\belowcaptionskip}{10pt}

        \subfigure[]
        {\includegraphics[width=0.6\textwidth]{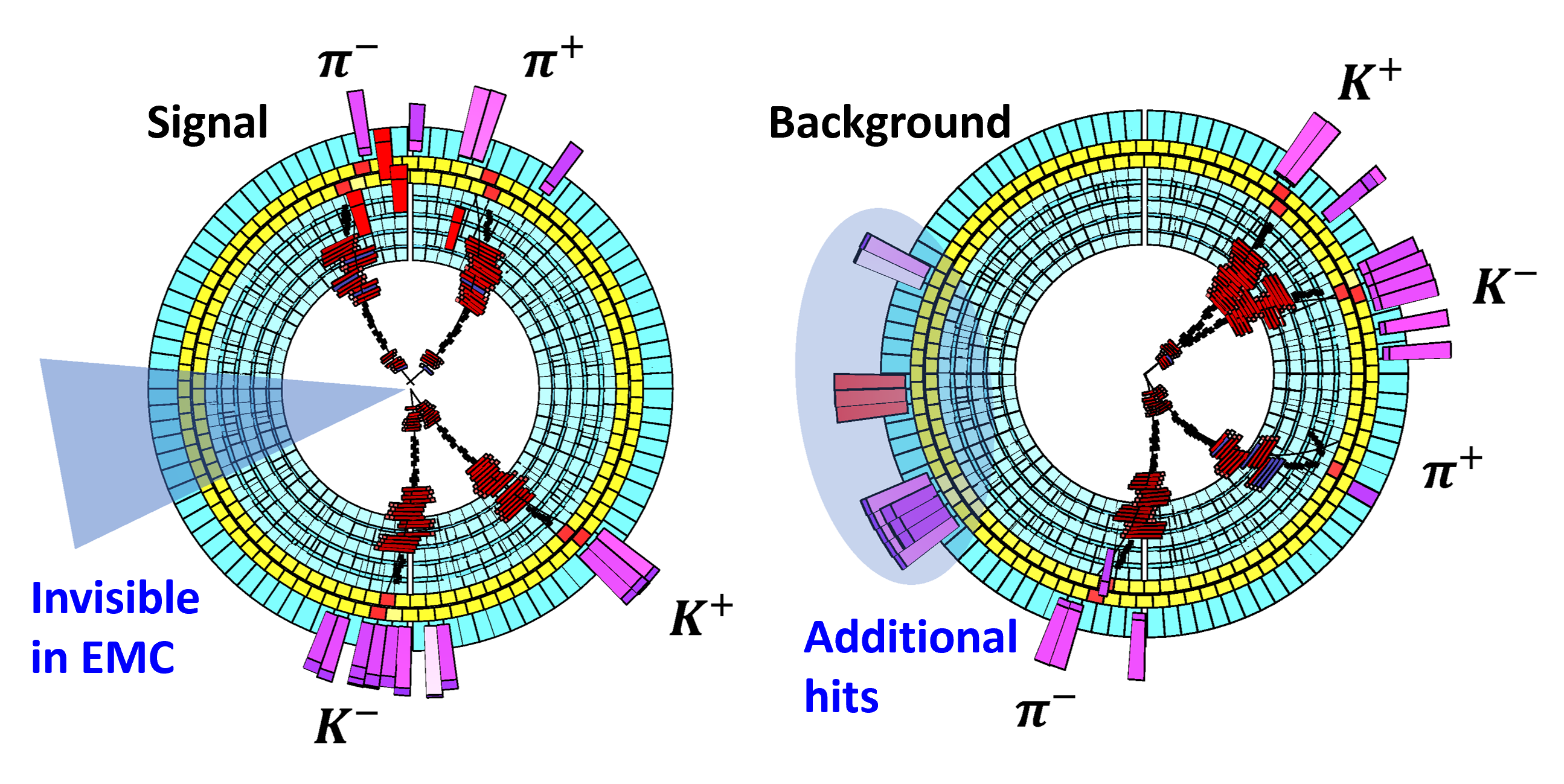}}
        \subfigure[]
        {\includegraphics[width=0.38\textwidth]{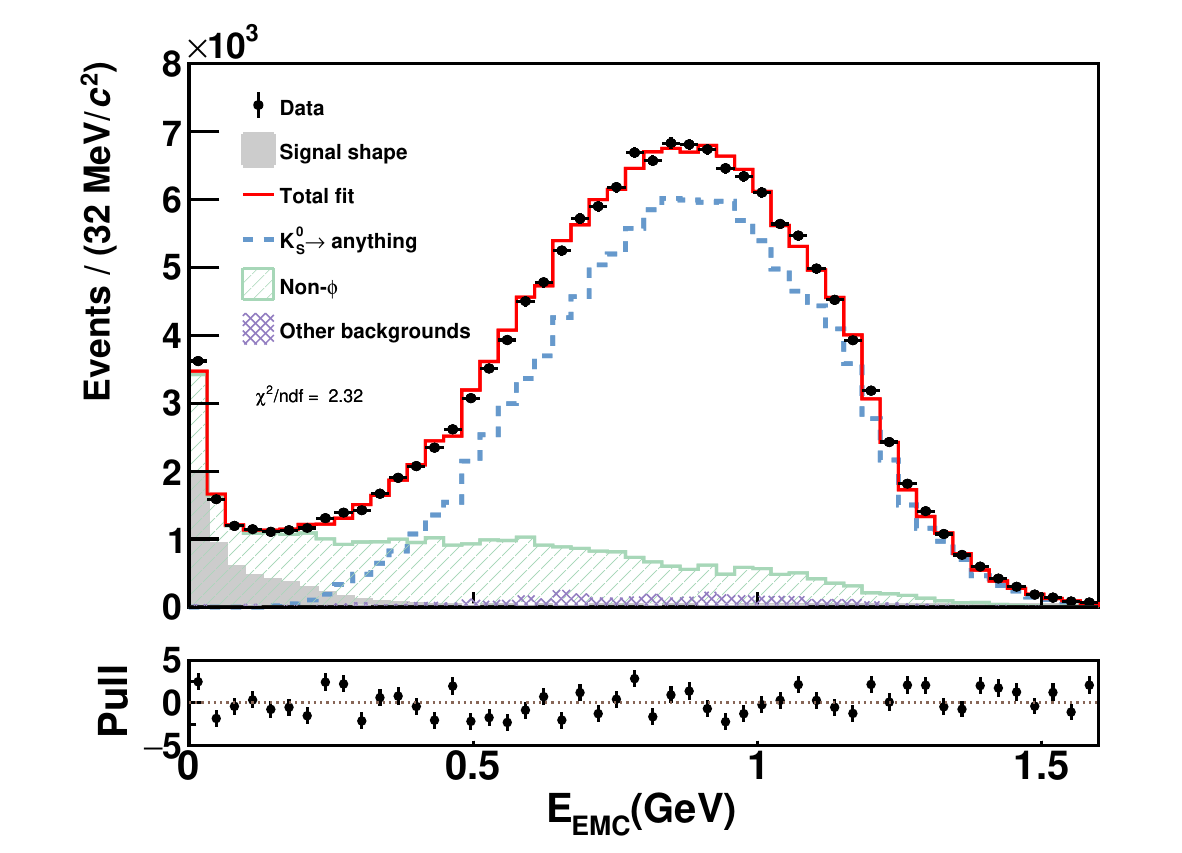}}\\
        
	\caption{
        (a) Illustration of the event displays for the $K^0_S \to \text{invisible}$ signal process (left) and a corresponding background process (right) obtained from Monte Carlo simulations. The EMC sub-detector is depicted by the light blue barrel, while the purple blocks indicate energy deposits within the EMC.
        (b) Distribution of the deposited energy for the selected $K^0_S \to \text{invisible}$ candidate events. The black data points represent the observed events, and the other lines or histograms correspond to various background and signal components.
        } 
	\label{fig:Ks}
\end{figure*}
\vspace{-0.0cm}

The first direct search for the invisible decay of $K^0_S$ is conducted at BESIII using a dataset of 10 billion $J/\psi$ events. The $K^0_S$ mesons are produced via the decay $J/\psi \to \phi K^0_S K^0_S$, which benefits from a relatively low background level due to the suppression of the decay $J/\psi \to \phi K^0_S K^0_L$ by C-parity conservation. In this analysis, the $\phi$ meson is reconstructed through the decay $\phi \to K^+ K^-$, one of the $K^0_S$ mesons is identified via $K^0_S \to \pi^+ \pi^-$, and the other $K^0_S$ meson is utilized to search for the invisible decay.
The invisible signal is identified using the Electron Magnetic Calorimeter (EMC) sub-detector. For signal events, no additional hits are recorded in the EMC, whereas SM background events typically exhibit some additional hits, as shown in Figure~\ref{fig:Ks} (a). Consequently, the deposited energy in the EMC is employed to extract the signal yield, as illustrated in Figure~\ref{fig:Ks} (b).
In the deposited energy distribution, the peak at zero corresponds to the background process $J/\psi \to K^+ K^- K^0_S K^0_L$ in the absence of a $\phi$ meson, which is characterized by the $\phi$ sideband region. No significant excess of events over the background is observed, and a UL on the BF of the invisible decay $K^0_S \to \text{invisible}$ is set to be $8.4 \times 10^{-4}$ at the 90\% C.L.~\cite{BESIII:2025kjj}. This result represents the first direct measurement of $K^0_S \to \text{invisible}$; however, the UL remains above the predictions from NP models, which require a larger data sample in future experiments.

\section{Search for dark baryon particles with $\Xi^-\to\pi^-\chi$}

Dark baryon is a hypothetical new particle that carries baryon number, and several hints may imply its existence. The energy densities of DM and baryonic matter in the universe are similar, with $\rho_{\rm{DM}} \sim 5.4\,\rho_{\rm{baryon}}$, suggesting a potential connection between their origins and motivating the existence of dark sector particles charged under baryon gauge symmetry. Additionally, in neutron lifetime measurements, the lifetime obtained from the beam method is longer than that from the storage method, which suggests an unknown BF for the decay $n\to\rm{dark~baryon}+X$ of approximately $1\%$.
Furthermore, considering the decay processes $B\to\rm{baryon}+\rm{dark~baryon}$ and the associated CP violation, the $B$-Mesogenesis mechanism can explain the asymmetry between visible matter and antimatter, as well as the origin and nature of DM~\cite{Alonso-Alvarez:2021oaj}. Naturally, dark baryons can interact with all SM quark flavors, not only in neutron or $B$ meson decays, providing an opportunity to search for dark baryon particles in hyperon decays at BESIII.

\begin{figure*}[htbp] \centering
	\setlength{\abovecaptionskip}{-1pt}
	\setlength{\belowcaptionskip}{10pt}

        \subfigure[]
        {\includegraphics[width=0.45\textwidth]{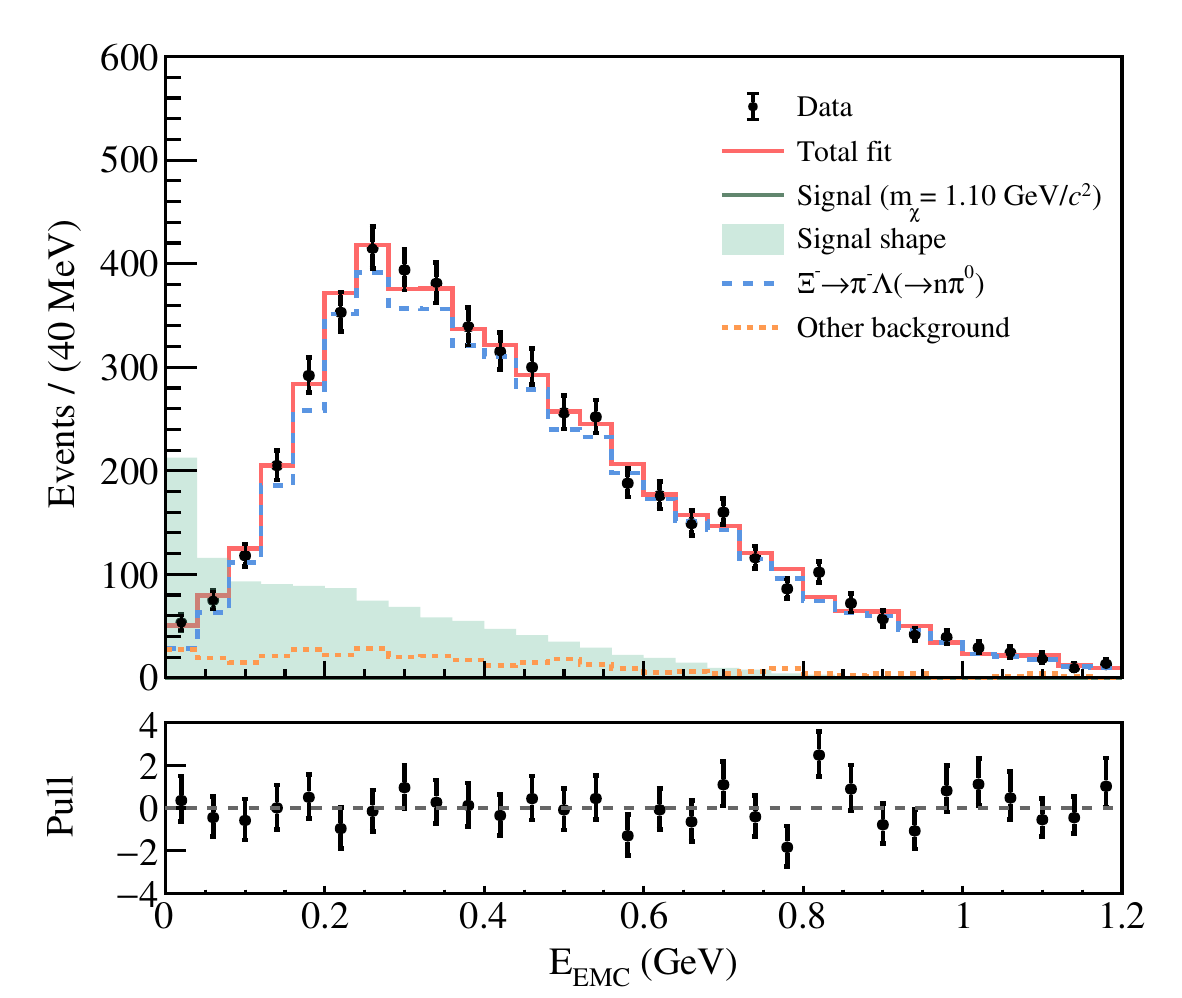}}
        \subfigure[]
        {\includegraphics[width=0.52\textwidth]{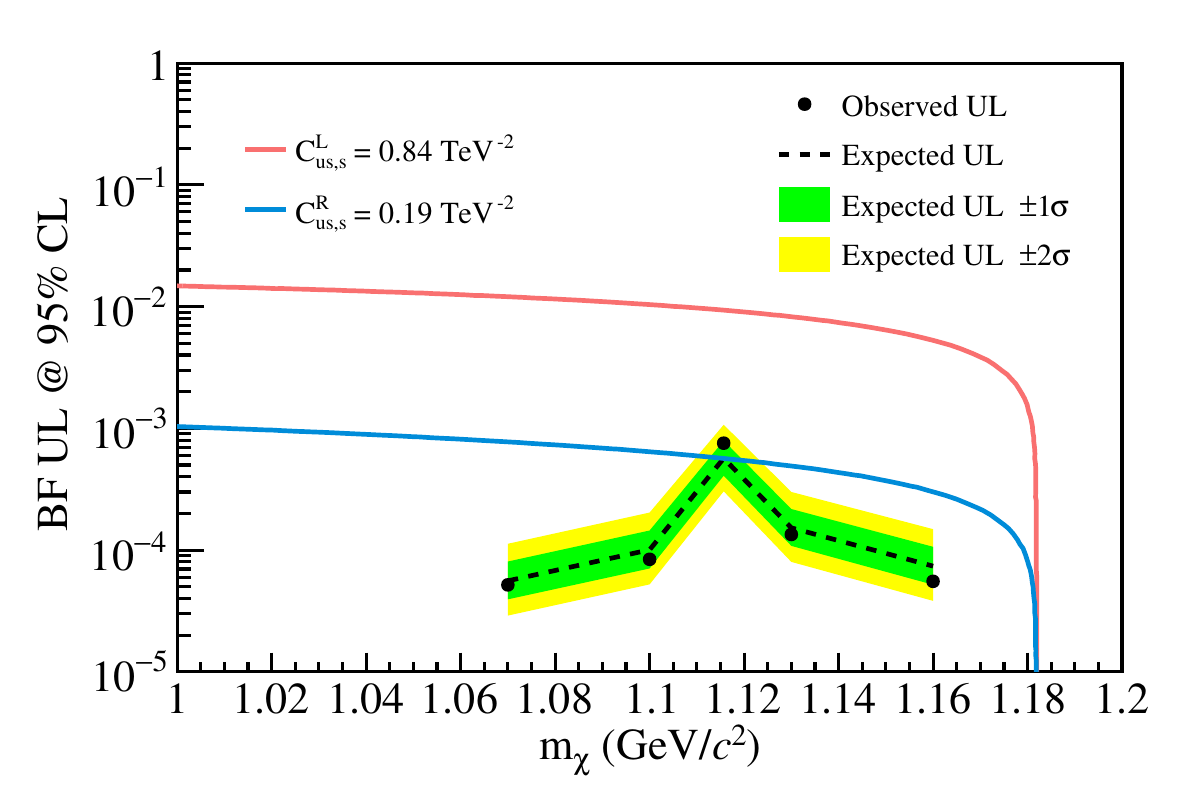}}\\
        
	\caption{
        (a) Distribution of the deposited energy for the selected $\Xi^- \to \pi^- \chi$ candidate events with $m_\chi = 1.10\,\rm{GeV}$. The black data points represent the observed events, while the other lines or histograms correspond to various background and signal components.
        (b) ULs on the BFs of $\Xi^- \to \pi^- \chi$ at the 95\% C.L.. The black points indicate the observed ULs obtained from the real data samples, the black dashed line and filled color regions represent the expected ULs, and the red and blue lines show the ULs recast from Ref.~\cite{Alonso-Alvarez:2021oaj}.
        } 
	\label{fig:Xi}
\end{figure*}

Searching for the dark baryon particle $\chi$ in the decay $\Xi^- \to \pi^- \chi$ is performed at BESIII for the first time~\cite{BESIII:2025sfl}. The analysis utilizes approximately $10^7$ $\Xi^-\bar{\Xi}^+$ pairs produced from $(10\,084 \pm 44) \times 10^6~J/\psi$ decays. The $\Xi^-$ candidate is identified by tagging a $\bar{\Xi}^+$ that decays to $\pi^+\bar{\Lambda}$ with $\bar{\Lambda} \to \bar{p}\pi^+$.
The analysis is conducted under dark baryon mass ($m_\chi$) hypotheses of $1.07\,\rm{GeV}$, $1.10\,\rm{GeV}$, $m_\Lambda$, $1.13\,\rm{GeV}$, and $1.16\,\rm{GeV}$, where $m_{\Lambda}$ is the known mass of the $\Lambda$ baryon. Constraints on the dark baryon mass are obtained by ensuring a good fit quality of a kinematic fit, which constrains the invariant mass recoil of the visible particles to $m_\chi$.
The invisible signal is identified using the EMC sub-detector, similar to the previous search for the invisible decay of $K^0_S$. For signal events, no additional hits are recorded in the EMC, whereas SM background events typically exhibit additional hits. The deposited energy in the EMC for $m_\chi = 1.10\,\rm{GeV}$ is shown in Figure~\ref{fig:Xi} (a), where no significant signals are observed. Similarly, no significant signals are observed for other $m_\chi$ hypotheses. The ULs on the BFs of $\Xi^- \to \pi^- \chi$ are presented in Figure~\ref{fig:Xi} (b), showing an improvement over constraints from the LHC (recast by Ref.~\cite{Alonso-Alvarez:2021oaj}) in the dark baryon model.

\vspace{-0.0cm}
\vspace{-0.0cm}

\section{Search for massless particles with $\Sigma^+\to p+\rm{invisible}$}

In the SM, the decay $\Sigma^+ \to p + \text{invisible}$ typically refers to $\Sigma^+ \to p \nu \bar{\nu}$, for which the BF is exceedingly small, primarily due to flavor-changing neutral currents and Glashow-Iliopoulos-Maiani mechanism suppression. However, the inclusion of other new invisible particles in the final state could potentially enhance its BF, such as through the emission of the QCD axion ($a$).
The QCD axion ($a$) was initially proposed by the Peccei-Quinn (PQ) mechanism as a solution to the strong CP problem and is also a viable candidate for cold dark matter. The mass of the QCD axion is inversely proportional to its decay constant $f_a$, given by
$m_a = 5.691(51)\,\mu\text{eV} \left( \frac{10^{12}\,\text{GeV}}{f_a} \right)$.
With decay constants $f_a \gg 10^6$ GeV, the axion mass $m_a$ remains below $\text{eV}$, rendering it effectively "massless" relative to the BESIII detector's resolution.
The QCD axion can couple to SM fermions via the interaction operator
$\mathcal{L}_{a-f} = \partial_{\mu} a \, \bar{f}_i \gamma^{\mu} \left( \frac{1}{F^V_{ij}} + \frac{\gamma^5}{F^A_{ij}} \right)$,
where $F^V_{ij}$ and $F^A_{ij}$ are the effective decay constants for the vector and axial coupling terms, respectively~\cite{MartinCamalich:2020dfe}. If the lepton U(1) charges are flavor non-universal, the QCD axion naturally acquires flavor-violating couplings, allowing for production via the decay $\Sigma^+ \to p + a$.
Furthermore, the massless dark photon, as introduced in the previous section, can be produced in the decay $\Sigma^+ \to p + \gamma'$, with the branching fraction bounded by $3.8 \times 10^{-5}$~\cite{MartinCamalich:2020dfe}.

\begin{figure*}[htbp] \centering
	\setlength{\abovecaptionskip}{-1pt}
	\setlength{\belowcaptionskip}{10pt}

        \subfigure[]
        {\includegraphics[width=0.45\textwidth]{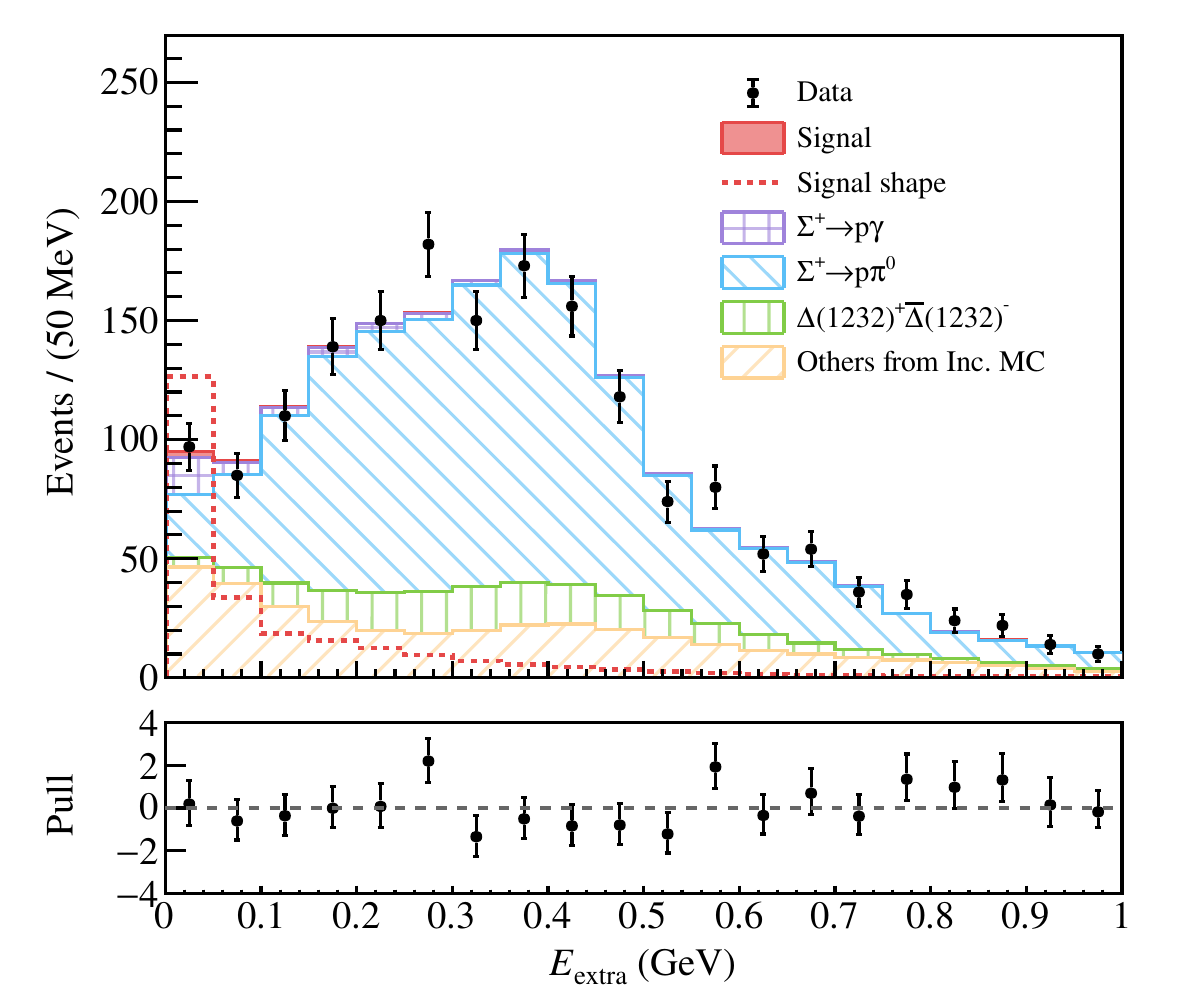}}
        \subfigure[]
        {\includegraphics[width=0.52\textwidth]{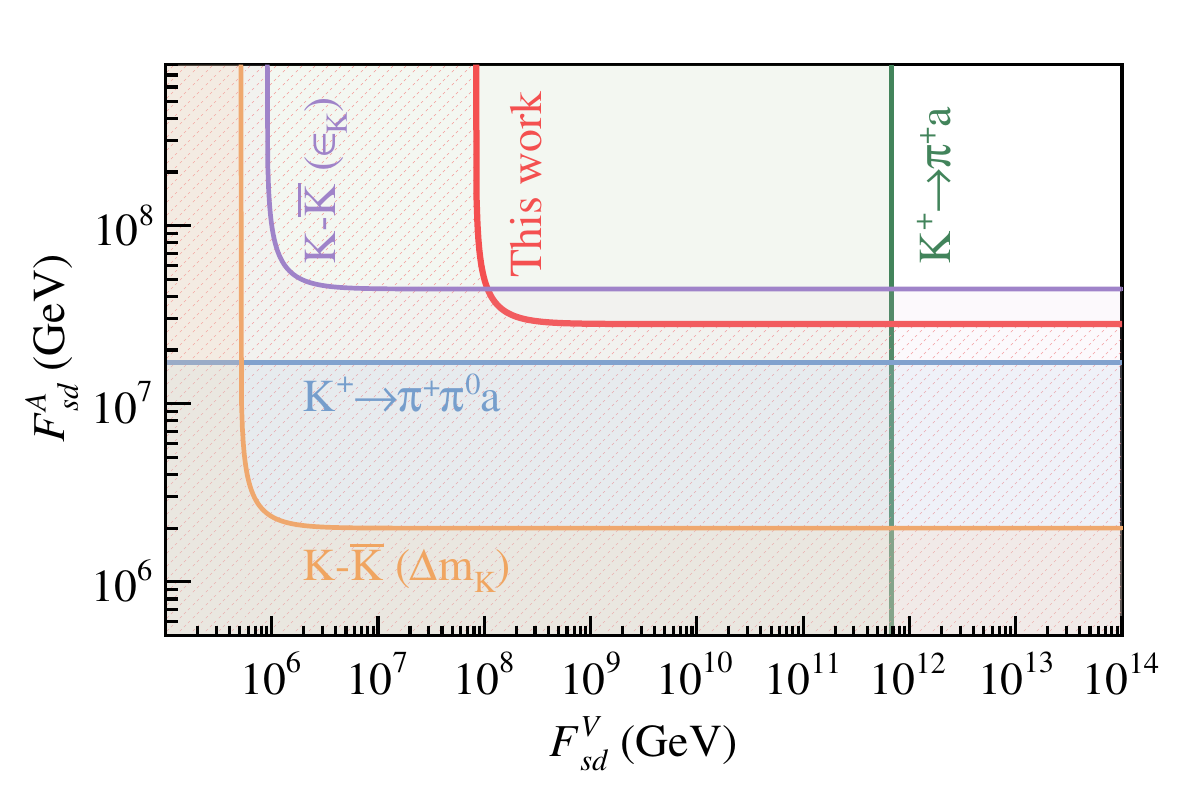}}
        
	\caption{
        (a) Distribution of the deposited energy for the selected $\Sigma^+ \to p +\rm{invisible}$ candidate events with $m_{\rm{invisible}} = 0$. The black data points represent the observed events, while the other lines or histograms correspond to various background and signal components.
        (b) The constraints on the effective decay constants for the vector coupling term and axial coupling term of QCD axion from different experiments.
        } 
	\label{fig:axion}
\end{figure*}

Using a dataset of $(10\,084 \pm 44) \times 10^6~J/\psi$ events decaying into $\Sigma^+\bar{\Sigma}^-$ pairs, we search for the massless particles $a$ or $\gamma'$ in the decay process $\Sigma^+ \to p + \text{invisible}$. A kinematic fit is applied to constrain the mass of the invisible particle to zero, and the signal is extracted from the energy spectrum of the additional shower detected in the EMC, as shown in Figure~\ref{fig:axion} (a). No significant signals are observed, and an UL on the BF $\mathcal{B}(\Sigma^+ \to p a)$ is set as $3.2 \times 10^{-5}$ at the 90\% C.L.. 
The constraints on the effective decay constants of the QCD axion are presented in Figure~\ref{fig:axion} (b). BESIII achieves a competitive limit on the axial-vector effective decay constant, establishing $F^A_{sd} > 2.8 \times 10^7$ GeV~\cite{BESIII:2023utd}.
\vspace{-0.0cm}
\vspace{-0.0cm}

\section{Search for axion-like particles in $J/\psi\to\gamma a$ with $a\to\gamma\gamma$}

The QCD axion, introduced in the previous section, is a compelling candidate for cold dark matter. However, no experimental signals of the QCD axion have been observed to date. A straightforward extension of the QCD axion model allows for arbitrary masses and coupling strengths, leading to the concept of axion-like particles (ALPs). ALPs interact with SM photons through the operator
$\mathcal{L} = -\frac{1}{4} g_{a\gamma\gamma} a F^{\mu\nu} \tilde{F}_{\mu\nu}$,
where $g_{a\gamma\gamma}$ denotes the coupling strength between the ALP and SM photons. ALPs can be produced from the decay of a heavy virtual photon via the process $\gamma^* \to a \gamma$ and subsequently decay through $a \to \gamma \gamma$ for detection.

\begin{figure*}[htbp] \centering
	\setlength{\abovecaptionskip}{-1pt}
	\setlength{\belowcaptionskip}{10pt}

        \subfigure[]
        {\includegraphics[width=0.49\textwidth]{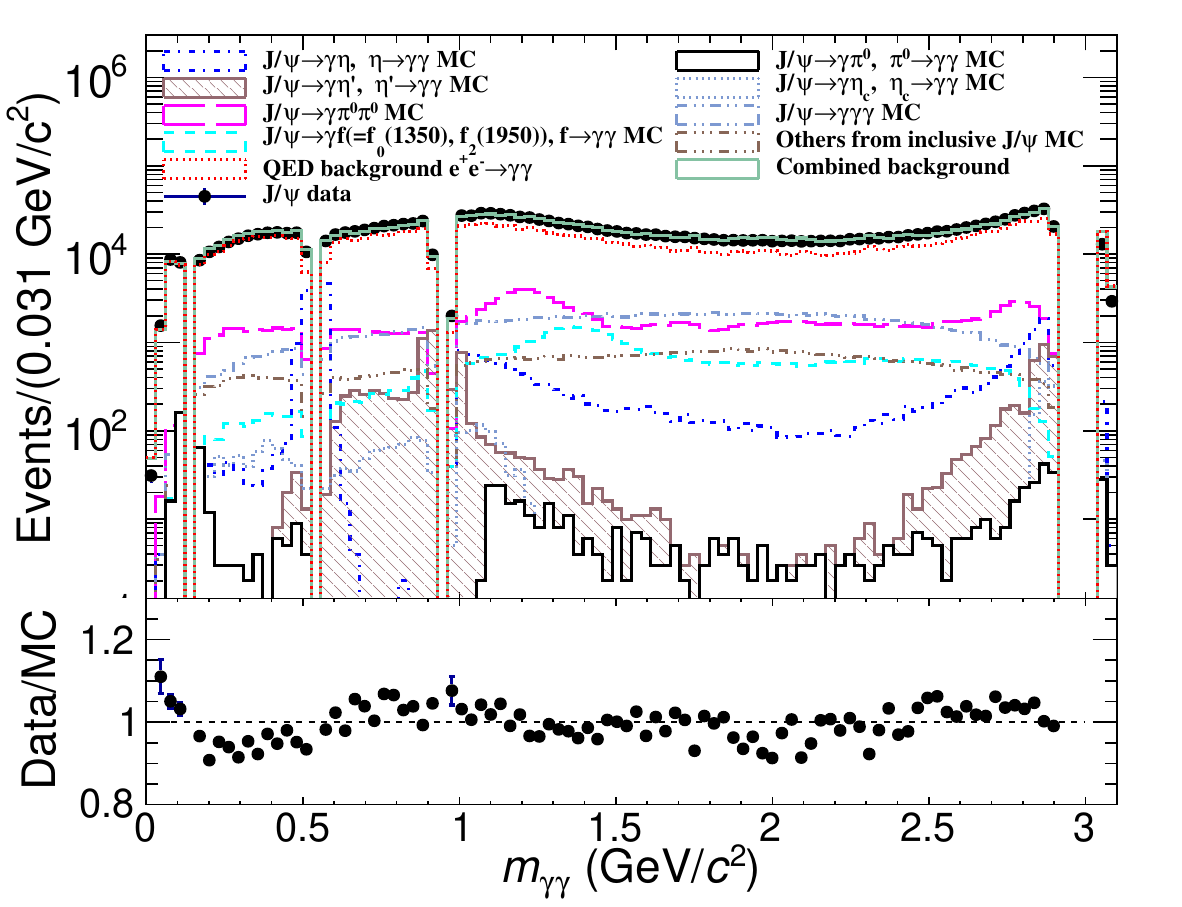}}
        \subfigure[]
        {\includegraphics[width=0.49\textwidth]{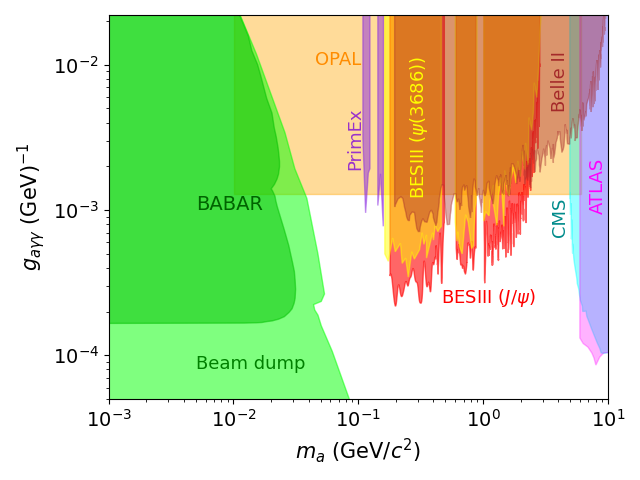}}
        
	\caption{
        (a) Distribution of the di-photon invariant mass in the ALP search. The black points represent the observed data, while the histogram shows contributions from various background processes. 
        (b) Constraints on the ALP-photon coupling. The red-filled regions indicate the excluded parameter space based on the latest BESIII results, whereas the other filled regions correspond to previously excluded parameter spaces.
        } 
	\label{fig:ALP}
\end{figure*}

At BESIII, the primary source of heavy photons is the $J/\psi$ events produced from $e^+e^-$ collisions. ALPs can be generated from the decay $J/\psi \to \gamma a$. The decay width of the ALP is given by $\Gamma_a = \frac{g^2_{a\gamma\gamma} m^3_a}{64\pi}$.
For a coupling strength of $g_{a\gamma\gamma} \sim 10^{-4}~\rm{GeV}^{-1}$ and an ALP mass $m_a \sim \rm{GeV}/c^2$, the ALP decays into two photons near the interaction point at BESIII.
Using a dataset of $(10\,084 \pm 44) \times 10^6~J/\psi$ events collected at BESIII, we search for ALPs in the process $J/\psi \to \gamma a \to \gamma \gamma \gamma$. Three photons are detected in the final state, and the ALP signal manifests as a peak in the $\gamma\gamma$ invariant mass distribution, as shown in Figure~\ref{fig:ALP} (a). To reduce background levels, events around known resonances such as $\pi^0$, $\eta$, $\eta'$, and $\eta_c$ in the $\gamma\gamma$ invariant mass distribution are excluded.
The ALP mass hypothesis is scanned range from $0.18$ to $2.85~\rm{GeV}/c^2$. The maximum global signal significance is found to be $1.6\sigma$ at $M_a = 2786~\rm{MeV}/c^{2}$, with no significant ALP signals observed. The corresponding ULs on the BFs for $J/\psi \to \gamma a \to \gamma \gamma \gamma$ are determined to range between $(3.6 - 53.1) \times 10^{-8}$ at the 95\% C.L.~\cite{BESIII:2024hdv}.

The BF of $J/\psi$ decaying into $\gamma a$ is sensitive to the coupling strength $g_{a\gamma\gamma}$, as described by~\cite{Merlo:2019anv}:
$\frac{\mathcal{B}(J/\psi \to \gamma a)}{\mathcal{B}(J/\psi \to e^+e^-)} = \frac{m^2_{J/\psi}}{32\pi \alpha} g^2_{a\gamma\gamma} \left( 1 - \frac{m^2_a}{m^2_{J/\psi}} \right)^3$.
In this analysis, the coupling between the ALP ($a$) and charm quarks is neglected. Assuming a BF of $\mathcal{B}(a \to \gamma \gamma) = 100\%$, the ULs on the ALP-photon coupling strength $g_{a\gamma\gamma}$ are set to be between $(2.2 - 101.8) \times 10^{-4}~\rm{GeV}^{-1}$ for ALP masses in the range $0.18 < m_a < 2.85~\rm{GeV}/c^2$, as shown in Figure~\ref{fig:ALP} (b). These represent the most stringent limits to date in this mass region.
Note that if the BF $\mathcal{B}(a \to \gamma \gamma)$ is less than $100\%$, the constraints from all experiments (not only BESIII) shown in Figure~\ref{fig:ALP} (b) would become less stringent.

\vspace{-0.0cm}
\vspace{-0.0cm}

\section*{Acknowledgements}
This work is supported by the National Key R\&D Program of China under Contracts Nos. 2023YFA1606000; National Natural Science Foundation of China~(Grant No. 12175321, 11975021).

\end{document}